\documentclass[preprint,aps,nofootinbib,showpacs]{revtex4}
\usepackage{amsmath,CJK}
\usepackage{amsmath,graphicx,color,epsfig}

\begin{document}
\begin{CJK*}{GBK}{Song}
\title{Study of $B\to K^{(*)} \ell^+\ell^-$ Decays in the Family Non-universal $Z'$
Models}\thanks{This work was supported in part by the National
Science Council of Taipei,  under Grant
No.~NSC~97-2112-M-008-002-MY3 and NCTS, and by the National Natural
Science Foundation of China under Grant No.~10735080, 11075168, and
10525523, and National Basic Research Program of China (973)
  No. 2010CB833000.}

\author{Cheng-Wei Chiang$^{b,c}$, LI Run-Hui$^{a,d}$, L\"U Cai-Dian$^a$}

\affiliation{
  \it $^a$ Institute of High Energy Physics and Theoretical Physics Center for Science Facilities,
  CAS, P.O. Box 918(4), Beijing 100049, China\\
  \it $^b$ Department of Physics and Center for Mathematics and
  Theoretical Physics, National Central University, Chungli  320,
   \\
  \it $^c$ Institute of Physics, Academia Sinica, Taipei  115, \\
  \it $^d$ School of Physics, Shandong University, Jinan 250100, China}

\begin{abstract}
  In a combined investigation of the $B\to K^{(*)}\ell^+\ell^-$ decays, constraints on the
  related couplings in family non-universal $Z^{\prime}$ models are derived.  We find that
  within the allowed parameter space, the recently observed forward-backward asymmetry in
  the $B\to K^*\ell^+\ell^-$ decay can be explained, by flipping the signs of the Wilson
  coefficients $C_9^{\rm eff}$ and $C_{10}$.  With the obtained constraints, we also
  calculate the branching ratio of the $B_s\to\mu^+\mu^-$ decay. The upper bound of our
  prediction is smaller by nearly an order than the upper bound given by CDF Collaboration recently.
\end{abstract}


 \maketitle
%==========================================================================================
\section{Introduction}
%==========================================================================================

The $B\to K^{(*)}\ell^+\ell^-$ decays play a very important role in heavy flavor physics.
 At the quark level, these decays involve the flavor-changing neutral current (FCNC) of the
  $b\to s$ transition, which is a purely quantum loop-mediated effect in the Standard Model (SM).
  Therefore, these decay modes have been proposed to test the SM predictions \cite{SMbasis}.
  In addition to the branching ratio, several observables of the $B\to K^*\ell^+\ell^-$ decay,
   such as the longitudinal polarization fraction, the forward-backward asymmetry ($A_{FB}$),
    the isospin symmetry, and the transverse asymmetry, have been proposed to probe possible
    new physics (NP) \cite{SM-NP}.  Various NP models thus have been scrutinized for their
    effects on these observables \cite{NPstudies}.

A few years ago, the forward-backward asymmetry of $B \to K^* \ell^+ \ell^-$ was first
 observed by the Belle Collaboration \cite{OldExp}.  The BaBar Collaboration also
 published its results in this channel earlier this year \cite{Babar:BtoK1,Barbar:BtoK2}.
 Recently, the Belle Collaboration updated its measurements in $B\to K^{(*)}\ell^+\ell^-$
 decays \cite{FBSexperiment}.  In these experiments, the forward-backward asymmetry is
  measured as a function of $q^2=M^2_{ll}c^2$, the invariant mass of the lepton pair.
  In comparison, BaBar only has two $q^2$ bins of data while Belle has six.  Their
  fitted $A_{FB}$ spectrum is generally higher than the SM expectation in all $q^2$ bins.
   This inspires us to do more investigations on these decays and see whether some NP
    model can better explain the experimental data.

In this paper, we consider a class of family non-universal
$Z^{\prime}$ models that induce FCNC's at tree level \cite{FCNCZpr}.
In such models, fermions in different families have different
couplings to the $Z'$ boson in the gauge basis.  After rotating to
the physical basis, off-diagonal couplings are generally produced,
inducing FCNC's at tree level.  These FCNC couplings are subject to
strong constraints from low-energy experiments.  Phenomenological
aspects of such models have been extensively analyzed by various
groups in recent years
\cite{ZprPheno1,ZprPheno2,ZprPheno3,ZprPheno4,ZprPheno5}.  In
particular, the possible $Z'$-$b$-$s$ coupling has received a lot of
attention because it may explain some of the puzzling $B$ physics
data.  Based on the previous analysis, we study whether the recently
observed $B \to K^{(*)} \ell^+ \ell^-$ data can be accommodated
within this model as well.

This paper is organized as follows.  In Sec.~\ref{sec:BtoKinSM}, we
first review the $B\to K^{(*)}\ell^+\ell^-$ decays in the SM.  In
the course, we define quantities relevant for the calculations, such
as form factors, effective Hamiltonian, explicit formulas of the
amplitudes, decays widths, and forward-backward asymmetries.  In
Sec.~\ref{sec:Zprime}, we describe the $Z'$ model with tree-level
FCNC's and deduce its effects on the $B\to K^{(*)}\ell^+\ell^-$
decays.  We then use the observables to constrain the model
parameters.  We find that the observed data in $B\to
K^{(*)}\ell^+\ell^-$ can be accommodated in such a $Z^{\prime}$
model.  We also predict the range of $Br(B_s \to \mu^+\mu^-)$ based
on the constrained parameter space.  Finally, we summarize our
findings in Sec.~\ref{sec:summary}.

%==========================================================================================
\section{$B\to K^{(*)}\ell^+\ell^-$ decays in the standard model} \label{sec:BtoKinSM}
%==========================================================================================
%----------------------------------------------------------------------
\subsection{Parametrization of the hadronic transitional matrix elements}
\label{sec:formfactor}
%----------------------------------------------------------------------

For the semileptonic decays investigated here, they involve hadronic matrix elements representing
 the $B \to K^{(*)}$ transitions.  Therefore, we first define the $B \to K$ form factors as
 follows:
\begin{eqnarray}
\langle K(p)|\bar s \gamma_{\mu} b|B(p_B)\rangle&=&
f_+(q^2)\left\{(p_B+p)_{\mu}-\frac{m_B^2-m_K^2}{q^2}q_{\mu}
\right\}+\frac{m_B^2-m_K^2}{q^2}f_0(q^2)q_{\mu}\;,\nonumber\\
\langle K(p)|\bar s \sigma_{\mu\nu}q^{\nu} b|B(p_B)\rangle&=&
i{(p_B+p)_{\mu}q^2-q_{\mu}(m_B^2-m_K^2)}\frac{f_T(q^2)}{m_B+m_K} ~,
\end{eqnarray}
where $q=p_B-p$ is the momentum transfer to the lepton pairs.  The $B\to K^*$ transitional form
 factors are defined as:
\begin{eqnarray}
  \langle {K^*}(p,\epsilon^*)|\bar q\gamma^{\mu}b|\bar B(p_B)\rangle
  &=&-\frac{2V(q^2)}{m_B+m_{K^*}}\epsilon^{\mu\nu\rho\sigma}
  \epsilon^*_{\nu}p_{B\rho}p_{\sigma}, \nonumber\\
  \langle {K^*}(p,\epsilon^*)|\bar q\gamma^{\mu}\gamma_5 b|\bar
  B(p_B)\rangle
  &=&2im_{K^*} A_0(q^2)\frac{\epsilon^*\cdot q}{q^2}q^{\mu}
  +i(m_B+m_{K^*})A_1(q^2)\left[\epsilon^*_{\mu}
    -\frac{\epsilon^*\cdot q}{q^2}q^{\mu} \right] \nonumber\\
  &&-iA_2(q^2)\frac{\epsilon^*\cdot q}{m_B+m_{K^*}}
  \left[ (p_B+p)^{\mu}-\frac{m_B^2-m_{K^*}^2}{q^2}q^{\mu} \right],\nonumber\\
  \langle {K^*}(p,\epsilon^*)|\bar q\sigma^{\mu\nu}q_{\nu}b|\bar
  B(p_B)\rangle
  &=&-2iT_1(q^2)\epsilon^{\mu\nu\rho\sigma}
  \epsilon^*_{\nu}p_{B\rho}p_{\sigma}, \nonumber\\
  \langle {K^*}(p,\epsilon^*)|\bar q\sigma^{\mu\nu}\gamma_5q_{\nu}b|\bar
  B(p_B)\rangle
  &=&T_2(q^2)\left[(m_B^2-m_{K^*}^2)\epsilon^{*\mu}
    -(\epsilon^*\cdot q)(p_B+p)^{\mu} \right]\nonumber\\
  &&+T_3(q^2)(\epsilon^*\cdot q)\left[
    q^{\mu}-\frac{q^2}{m_B^2-m_{K^*}^2}(p_B+p)^{\mu}\right] ~.
  \label{eq:BtoVformfactors}
\end{eqnarray}
In the calculations of the semileptonic decays, we need the $q^2$ dependence in the form factors.
For $B\to K^*$ transitions, we adopt the dipole model parametrization for the form factors:
\begin{eqnarray}
  F(q^2)=\frac{F(0)}{1-a(q^2/m_B^2)+b(q^2/m_B^2)^2},
\end{eqnarray}
where $a$ and $b$ are parameters to be determined.  We calculate the form factors in the PQCD
approach \cite{PQCD} near the $q^2=0$ region, where the $K^*$ meson recoils very fast, and
determine their values at some points.  Then we extrapolate our results to the entire kinematic
regime through fitting.  Our results in the PQCD approach as well as those obtained using the
QCD sum rules (QCDSR) \cite{formfactorsQCDSR} are listed in Table~\ref{tab:formfactors}.
In our calculations, we will mainly use the PQCD results.  The QCDSR results are included only
as a comparison because we do not have the explicit errors on the QCDSR results .

For the form factors of $B\to K$ transition, we adopt a different
parametrization:
\begin{eqnarray}
  F(q^2)=F(0) \exp \left[
c_1(q^2/m_B^2)+c_2(q^2/m_B^2)^2+c_3(q^2/m_B^2)^3 \right] ~,
\label{eq:LCSRpara}
\end{eqnarray}
because the authors of Ref.~\cite{BtoKbyAli} find that in their fitting the extrapolation of
the dipole parametrization to maximum $q^2$ is prone to reach a serious singularity below the
physical cut starting at $q^2=m_B^2$.  The values of the parameters in the $B\to K$ form factors
\cite{BtoKbyAli} are listed in Table~\ref{tab:formfactorsBtoK}.

%%%%%%%%%%%%%%%%%%%%%%%%%%%%%%%%%%%%%%%%%%%%%%%%%%%%%%%%%%%%%%%%%%%%%
%%%%   form factors for B to K* in PQCD and QCD sum rules
%%%%%%%%%%%%%%%%%%%%%%%%%%%%%%%%%%%%%%%%%%%%%%%%%%%%%%%%%%%%%%%%%%%%%

\begin{table}
 \caption{$B\to K^*$ form factors in PQCD approach and QCD sum rules (QCDSR).}
 \label{tab:formfactors}
\begin{center}
\begin{tabular}{c|cc||c|cc}
\hline\hline
 \ \ \ $$    &PQCD     &QCDSR \cite{formfactorsQCDSR}   &$$    &PQCD     & QCDSR \cite{formfactorsQCDSR}\\
\hline
 \ \ \ $V(0)$     &$0.26$     &$0.458$        &$T_1(0)$     &$0.23$     &$0.379$ \\
 \ \ \ $a(V)$     &$1.75$     &$1.55$        &$a(T_1)$     &$1.70$     &$1.59$ \\
 \ \ \ $b(V)$     &$0.68$     &$0.575$        &$b(T_1)$     &$0.63$     &$0.615$ \\
 \hline
 \ \ \ $A_0(0)$     &$0.30$     &$0.470$        &$T_2(0)$     &$0.23$     &$0.379$ \\
 \ \ \ $a(A_0)$     &$1.72$     &$1.55$        &$a(T_2)$     &$0.71$     &$0.49$ \\
 \ \ \ $b(A_0)$     &$0.62$     &$0.680$        &$b(T_2)$     &$-0.19$     &$-0.241$ \\
 \hline
 \ \ \ $A_1(0)$     &$0.19$     &$0.337$        &$T_3(0)$     &$0.20$     &$0.261$ \\
 \ \ \ $a(A_1)$     &$0.79$     &$0.60$        &$a(T_3)$     &$1.58$     &$1.20$ \\
 \ \ \ $b(A_1)$     &$-0.09$    &$-0.023$       &$b(T_3)$     &$0.49$     &$0.098$ \\
 \hline
 \ \ \ $A_2(0)$     &$$     &$0.283$        &$$     &$$     &$$ \\
 \ \ \ $a(A_2)$     &$$     &$1.18$        &$$     &$$     &$$ \\
 \ \ \ $b(A_2)$     &$$     &$0.281$        &$$     &$$     &$$ \\
\hline\hline
\end{tabular}
\end{center}
\end{table}

%%%%%%%%%%%%%%%%%%%%%%%%%%%%%%%%%%%%%%%%%%%%%%%%%%%%%%%%%%%%%%%%%%%%%
%%%%   form factors for B to K sum rules
%%%%%%%%%%%%%%%%%%%%%%%%%%%%%%%%%%%%%%%%%%%%%%%%%%%%%%%%%%%%%%%%%%%%%

\begin{table}
\caption{$B\to K$ form factors in light cone sum rules with the
parametrization, Eq.~(\ref{eq:LCSRpara}).}
 \label{tab:formfactorsBtoK}
\begin{center}
\begin{tabular}{c|cccc}
\hline\hline
 \ \ \        & $F(0)$    &$c_1$           &$c_2$          &$c_3$\\
\hline
 \ \ \ $f_+(q^2)$      &$0.319$     &$1.465$        &$0.372$        &$0.782$  \\
 \ \ \ $f_0(q^2)$     &$0.319$     &$0.633$        &$-0.095$       &$0.591$   \\
 \ \ \ $f_T(q^2)$     &$0.355$     &$1.478$        &$0.373$        &$0.700$  \\
\hline\hline
\end{tabular}
\end{center}
\end{table}

%----------------------------------------------------------------------
\subsection{Effective Hamiltonian and decay amplitudes}
\label{sec:Hamiltonian}
%----------------------------------------------------------------------

At the quark level, the $B\to K^{(*)}\ell^+\ell^-$ decays are dominated by the
$b\to s\ell^+\ell^-$ transition, the Hamiltonian for which is given by
\begin{eqnarray}
{\cal H}_{\rm{eff}} =
-\frac{G_F}{\sqrt{2}} V_{tb} V^*_{ts} \sum_{i=1}^{10} C_i(\mu) O_i(\mu) ~,
\label{eq:Hamiltonian}
\end{eqnarray}
where $V_{tb}$ and $V_{ts}$ are the Cabibbo-Kobayashi-Maskawa (CKM) matrix elements and
 $C_i(\mu)$ are the Wilson coefficients evaluated at the scale $\mu$.
 The local operators $O_i(\mu)$ are given by \cite{Buchalla:1995vs}
\begin{eqnarray}
 O_1&=&(\bar s_{\alpha}c_{\alpha})_{V-A}(\bar
 c_{\beta}b_{\beta})_{V-A},\;\;
 O_2=(\bar
 s_{\alpha}c_{\beta})_{V-A}(\bar
 c_{\beta}b_{\alpha})_{V-A},\nonumber\\
 %-------------------------------------------------
 O_3&=&(\bar s_{\alpha}b_{\alpha})_{V-A}\sum_q(\bar
 q_{\beta}q_{\beta})_{V-A},\;\;
 O_4=(\bar s_{\alpha}b_{\beta})_{V-A}\sum_q(\bar
 q_{\beta}q_{\alpha})_{V-A},\nonumber\\
 %-------------------------------------------------
 O_5&=&(\bar s_{\alpha}b_{\alpha})_{V-A}\sum_q(\bar
 q_{\beta}q_{\beta})_{V+A},\;\;
 O_6=(\bar s_{\alpha}b_{\beta})_{V-A}\sum_q(\bar
 q_{\beta}q_{\alpha})_{V+A},\nonumber\\
 %-------------------------------------------------
 O_7&=&\frac{e m_b}{8\pi^2}\bar
 s\sigma^{\mu\nu}(1+\gamma_5)bF_{\mu\nu}+\frac{e m_s}{8\pi^2}\bar
 s\sigma^{\mu\nu}(1-\gamma_5)bF_{\mu\nu},\nonumber\\
 %------------------------------------------------------
 O_9&=&\frac{\alpha_{\rm{em}}}{2\pi}(\bar \ell\gamma_{\mu}\ell)(\bar
 s\gamma^{\mu}(1-\gamma_5)b),\;\;
 O_{10}=\frac{\alpha_{\rm{em}}}{2\pi}(\bar \ell\gamma_{\mu}\gamma_5\ell)(\bar
 s\gamma^{\mu}(1-\gamma_5)b) ~,
\label{eq:operators}
\end{eqnarray}
where $\alpha$ and $\beta$ are color indices, $q=u,d,s,c$, $(\bar q_1q_2)_{V-A}(\bar q_3 q_4)_{V-A}
\equiv [ \bar q_1
\gamma^{\mu}(1-\gamma_5)q_2 ] [\bar q_3\gamma_{\mu}(1-\gamma)q_4 ]$,
and $(\bar q_1q_2)_{V-A}(\bar q_3 q_4)_{V+A} \equiv [ \bar q_1
\gamma^{\mu} (1-\gamma_5) q_2][ \bar q_3\gamma_{\mu}(1+\gamma_5)q_4
]$.

With the above Hamiltonian, the amplitude of $b\to s\ell^+\ell^-$ transition can be written as
\begin{eqnarray}
  && {\cal A}(b\to s\ell^+
  \ell^-) \nonumber \\
  && \qquad = \frac{G_F}{2\sqrt{2}}\frac{\alpha_{\rm{em}}}{\pi}V_{tb}V^*_{ts}\bigg\{
  C_9^{\rm{eff}}(q^2)
  [\bar s \gamma_{\mu}(1-\gamma_5)b][\bar \ell\gamma^{\mu}\ell]
  + C_{10}[\bar s\gamma_{\mu}(1-\gamma_5)b]
  [\bar \ell\gamma^{\mu}\gamma_5\ell]\nonumber\\
  && \qquad - 2m_bC_7^{\rm{eff}}\big[\bar s i\sigma_{\mu\nu}\frac{q^{\nu}}{q^2}
  (1+\gamma_5)b\big][\bar \ell\gamma^{\mu}\ell]- 2m_sC_7^{\rm{eff}}\big[\bar s i\sigma_{\mu\nu}
  \frac{q^{\nu}}{q^2}
  (1-\gamma_5)b\big ][\bar \ell\gamma^{\mu}\ell] \bigg\} ~,
\label{eq:Ampbtos}
\end{eqnarray}
where $m_b$ is the $b$ quark mass in the $\overline{\mbox{MS}}$ scheme.  The Wilson coefficients
$C_7^{\rm{eff}} = C_7 - C_5/3 - C_6$ and $C_9^{\rm{eff}}$ contain both the long-distance and
short-distance contributions:
\begin{eqnarray}
  C_9^{\rm{eff}}(q^2)&=&C_9(\mu)+Y_{\rm{pert}}(q^2)+Y_{\rm{LD}}(q^2) ~.
\end{eqnarray}
Here $Y_{\rm{pert}}$ represents the perturbative contribution, and $Y_{\rm{LD}}$ is the
long-distance part containing contributions from the resonant states and can be excluded by
experimental analysis.  Thus we will not include $Y_{\rm LD}$ in our calculation, and
\begin{eqnarray}
 C_9^{\rm{eff}}(q^2)&=&C_9(\mu)+Y_{\rm{pert}}(q^2) ~,
\label{eq:C9}
\end{eqnarray}
with the detailed form of $Y_{\rm{pert}}$ given in Ref.~\cite{ypert}.

The $B\to K^*\ell^+\ell^-$ decay is more complicated because of its polarization structures in
the final state.  We will use the helicity basis.  By re-expressing the metric tensor
\begin{eqnarray}
 g_{\mu\nu}=
-\sum_\lambda\epsilon_\mu(\lambda) \epsilon^*_\nu(\lambda)
+\frac{q_\mu q_\nu}{q^2} ~,
\label{eq:replacement}
\end{eqnarray}
we can decompose the amplitude ${\cal A}(\bar B\to K^*\ell^+\ell^-)$ into Lorentz-invariant
leptonic part $L(L/R,\lambda)$ and hadronic part $H(L/R,\lambda)$:
\begin{eqnarray}
  {\cal A}(\bar B\to \bar K^*\ell^+\ell^-)&=&
  {L}_\mu(L) {H}_\nu(L) g^{\mu\nu}+{L}_\mu(R) {H}_\nu(R) g^{\mu\nu}
  \nonumber\\
  &=&-\sum_{\lambda}{L}(L,\lambda) {H}(L,\lambda)
  -\sum_{\lambda}{L}(R,\lambda) {H}(R,\lambda) ~.
\end{eqnarray}
The details have been given in Appendix C of Ref.~\cite{BtoK1}.  The explicit
formulas of the functions $L(L/R,\lambda)$ and $H(L/R,\lambda)$ are
listed in Appendix~\ref{appendix:LH}.

%-------------------------------------------------------------------------
\subsection{The decay widths and branching ratios}
\label{sec:brs}
%-------------------------------------------------------------------------

With the form factors given in Sec.~\ref{sec:formfactor} and
Eq.~(\ref{eq:Ampbtos}), we obtain the dilepton spectrum of $B\to K
\ell^+\ell^-$ as
\begin{eqnarray}
\frac{d\Gamma_i(B\to
K\ell^+\ell^-)}{dq^2}&=&
\frac{G_F^2|V_{tb}|^2|V_{ts}^*|^2\alpha_{em}^2\lambda^{3/2}}{1536 \pi^5 m_B^3}
\left\{|C_{10}f_+(q^2)|^2 \right. \nonumber \\
&& \qquad + \left.
\left| C_9^{\rm{eff}}f_+(q^2)+\frac{2C_7^{\rm{eff}}(m_b+m_s)}{m_B+m_K}f_T(q^2) \right|^2\right\} ~,
\label{eq:gammaBtoK}
\end{eqnarray}
where
\begin{eqnarray}
\lambda=(m_{K^*}^2+m_B^2-q^2)^2-4m_B^2m_{K^*}^2 =
(m_B^2-m_{K^*}^2-q^2)^2-4m_{K^*}^2q^2 ~.
\label{eq:lambda}
\end{eqnarray}

For the $B\to K^*\ell^+\ell^-$ decay, we define the direction opposite to the momentum of $K^*$
meson in the rest frame of the $B$ meson as the $+z$ direction.  In the center-of-mass (CM) frame
 of $\ell^+\ell^-$, $\theta_1$ is defined as the angle between the $z$ axis and the momentum of
 $\ell^-$.  In the experiment, the $K^*$ meson usually decays to the $K\pi$ final state.  We define
 the angle between the decay plane $K^*\to K\pi$ and the plane determined by $\ell^+\ell^-$ as $\phi$.
 Combining the leptonic amplitudes, the hadronic amplitudes, and the phase space all together,
  the partial decay width of $B\to K^*\ell^+\ell^-$ is given by
\begin{eqnarray}
  && d\Gamma_i(\bar B\to \bar K^*\ell^+\ell^-)
  =\frac{\sqrt{\lambda}}{1024 \pi^4 m_B^3}
  d\cos\theta_1 d\phi dq^2 |{\cal A}_i(B\to
  K^*\ell^+\ell^-)|^2
  \nonumber\\
  &&=\frac{\sqrt{\lambda}}{1024 \pi^4 m_B^3}
  d\cos\theta_1 d\phi dq^2 (| L(L,i) H(L,i)|^2+| L(R,i) H(R,i)|^2) ~,
  \label{eq:Pdecaywith}
\end{eqnarray}
where $i=0,+$ or $-$ denotes the three different polarizations of the $K^*$.

After integrating out $\theta_1$ and $\phi$ in Eq.~(\ref{eq:Pdecaywith}), one obtains the dilepton
 spectrum of $B\to {K^*}\ell^+\ell^-$ decay as:
\begin{eqnarray}
  \frac{d\Gamma_i(B\to K^*\ell^+\ell^-)}{dq^2}=\frac{\sqrt{\lambda}q^2}{96 \pi^3 m_B^3}
  \bigg[|H(L,i)|^2+|H(R,i)|^2\bigg] ~.
\label{eq:Gammaq2}
\end{eqnarray}
%%----------------------------------------------------------------
%\begin{figure*}[htb]
%\includegraphics[scale=0.4]{dBr.eps}
%\caption{The dilepton spectrum with form factors gained in QCD sum
%rules(black solid line) and in PQCD approach(red dotted line).}
%\label{fig:dilepton}
%\end{figure*}
%%----------------------------------------------------------------
%The dilepton spectrum we obtained is shown in
%Fig.~\ref{fig:dilepton}, and the branching ratios in experiment and
%in our results(only central value) are
%\begin{eqnarray}
%Br(\bar B \to \bar K^*\ell^+\ell^-)&=&13.9\times 10^{-7}\,\,\,\,\,\,\,\,\,\,\mbox{in PQCD}\;,\label{eq:BrinPQCD}\\
%Br(\bar B \to \bar K^*\ell^+\ell^-)&=&15.6\times 10^{-7}\,\,\,\,\,\,\,\,\,\,\mbox{in QCDSR}\;,\label{eq:BrinQCDSR}\\
%Br(\bar B \to \bar K^*\ell^+\ell^-)&=&(10.7_{-1.0}^{+1.1}\pm0.9)\times 10^{-7}\,\,\,\,\,\,\mbox{in experiment\cite{FBSexperiment}}\;,\\
%Br(\bar B \to \bar K\ell^+\ell^-)&=&(4.8_{-0.4}^{+0.5}\pm0.3)\times 10^{-7}\,\,\,\,\,\,\,\,\,\,\mbox{in experiment}\;,\label{eq:BrinPQCD}\\
%\end{eqnarray}

%****************************************************************************
\begin{figure*}[htb]
  \includegraphics[width=7.5cm,height=8cm]{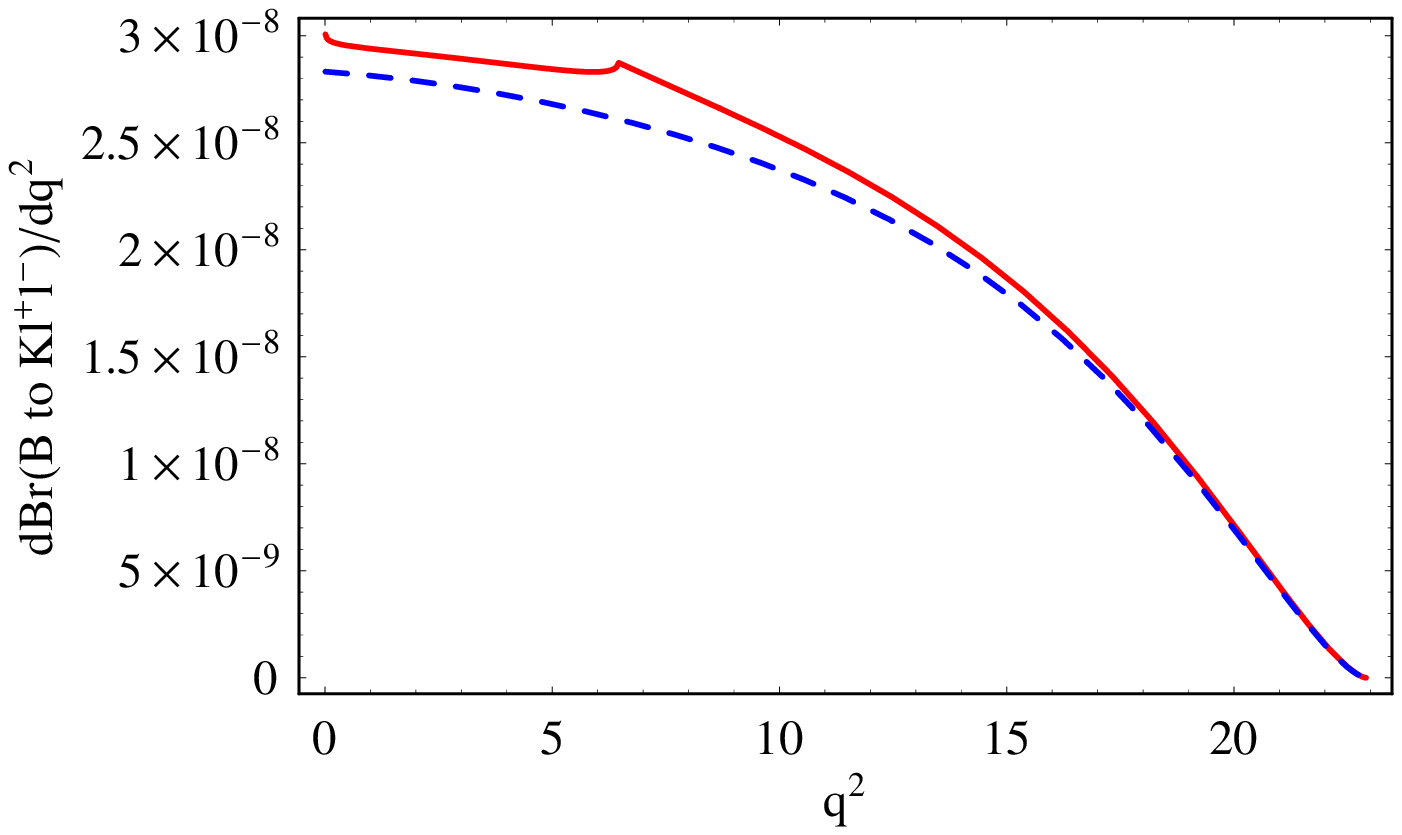}
  \hspace{1cm}
  \includegraphics[width=7.5cm,height=8cm]{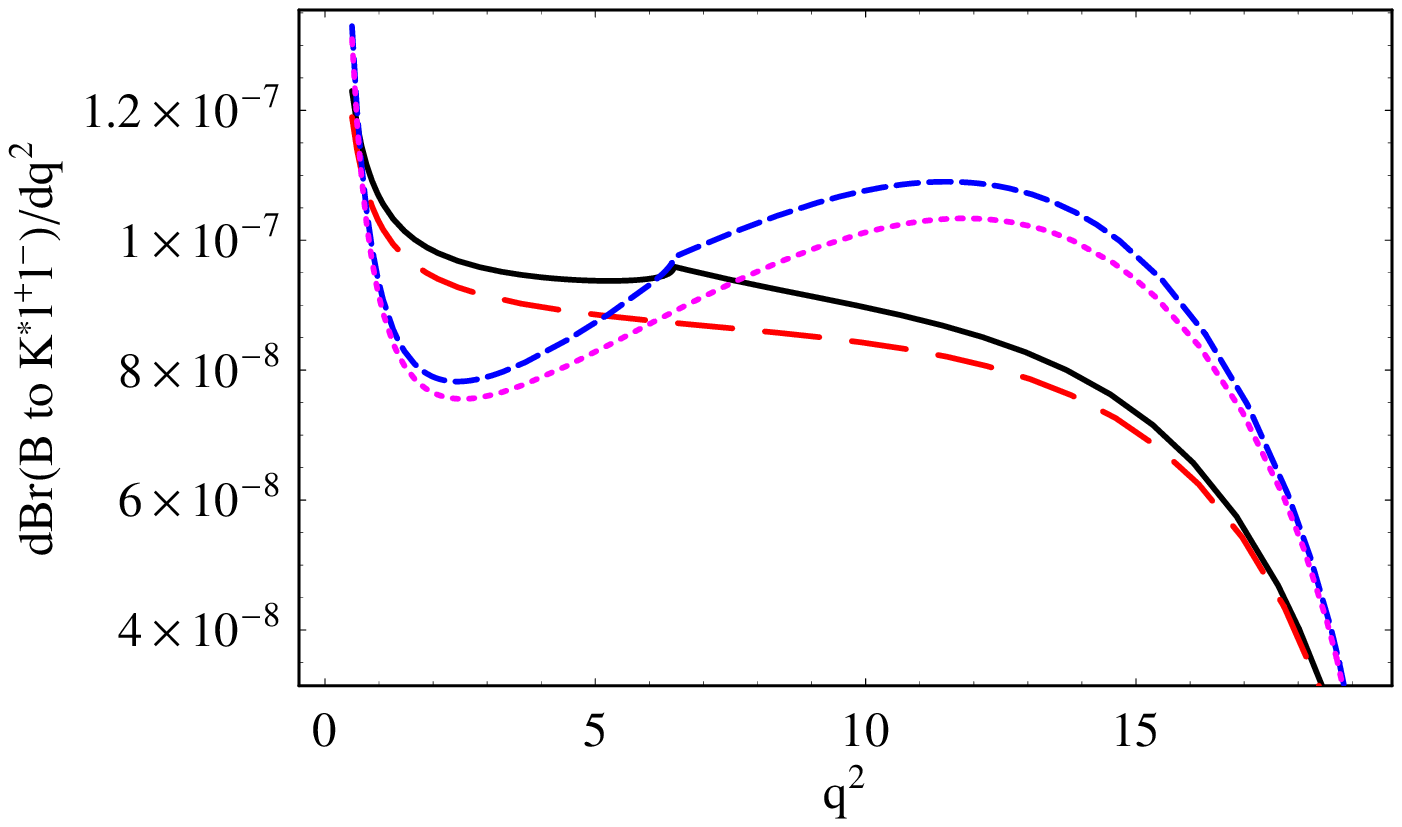}
 \vspace{-2cm}
 \caption{$q^2$-dependence of the branching ratios of
$B\to K\ell^+\ell^-$ (left plot) and $B\to K^*\ell^+\ell^-$ (right plot) decays.
In the left plot, the red solid (blue dashed) curve stands for the
dilepton spectrum with (without) the $Y_{\rm{pert}}(q^2)$ part
included in $C_9^{\rm{eff}}$.  The right plot shows the spectrum
predicted in PQCD and QCDSR with and without the
$Y_{\rm{pert}}(q^2)$ part in Eq.~(\ref{eq:C9}). The black solid (red
long dashed) curve is the PQCD results with (without)
$Y_{\rm{pert}}(q^2)$ and the blue short dashed (pink dotted) curve
is the QCDSR result with (without) $Y_{\rm{pert}}(q^2)$. In the
curves where $Y_{\rm{pert}}(q^2)$ is included, a kink shows up
because it is a piecewise function.} \label{Fig:BtoK}
\end{figure*}
%*******************************************************************************

In Sec.~\ref{sec:Hamiltonian}, one can find that among the Wilson coefficients only $C_9^{eff}$
has the $q^2$ dependence.  The dilepton spectra of $B\to K^{(*)}\ell^+\ell^-$ decays are shown in
Fig.~\ref{Fig:BtoK}, with and without $Y_{\rm{pert}}(q^2)$ in $C_9^{\rm{eff}}$ being included.
 After further integrating out the $q^2$ dependence, we obtain the total branching ratios:
\begin{eqnarray}
  Br(B\to K\ell^+\ell^-)=\bigg\{
 \begin{array}{cc}(4.70_{-0.71}^{+1.29})\times 10^{-7} \,\,\,\,\,\,\mbox{($q^2$ part in $C_9^{\rm{eff}}$ included),} \\
 (4.45_{-0.67}^{+1.22})\times 10^{-7} \,\,\,\,\,\,\mbox{($q^2$ part in $C_9^{\rm{eff}}$ excluded),}\\
 \end{array}
\label{eq:NBrBtoK}\\
%-------------------------------------------------------------------------------------------------------------------------
Br(B\to K^*\ell^+\ell^-)=\bigg\{
 \begin{array}{cc}(16.5_{-5.7}^{+7.8})\times 10^{-7} \,\,\,\,\,\,\mbox{($q^2$ part in $C_9^{\rm{eff}}$ included),} \\
 (15.8_{-5.5}^{+7.5})\times 10^{-7} \,\,\,\,\,\,\mbox{($q^2$ part in $C_9^{\rm{eff}}$ excluded).}\\
 \end{array}
\label{eq:NBrBtoKst}
\end{eqnarray}
These predictions are to be compared with the experimental results \cite{FBSexperiment}:
\begin{eqnarray}
Br(B \to K\ell^+\ell^-)&=&(4.8_{-0.4}^{+0.5}\pm0.3)\times
10^{-7} ~,
\nonumber \\
Br(B \to K^*\ell^+\ell^-)&=&(10.7_{-1.0}^{+1.1}\pm0.9)\times
10^{-7} ~.
\end{eqnarray}
From Fig.~\ref{Fig:BtoK}, Eq.~(\ref{eq:NBrBtoK}) and Eq.~(\ref{eq:NBrBtoKst}), one finds that the
$Y_{\rm{pert}}(q^2)$ piece in $C_9^{\rm{eff}}$ has a small effect on the branching ratios in
comparison with other uncertainties.  To simplify the notation, we define
$C_9^{\prime}\equiv Y_{\rm{pert}}(q^2)$, and thus $C_9^{\rm{eff}}=C_9+C_9^{\prime}$.
The differential branching ratio of $B\to K\ell^+\ell^-$ is then decomposed into the following
form
\begin{eqnarray}
&& \frac{dBr(B\to K\ell^+\ell^-)}{dq^2}
= |C_{10}|^2 B_1^{\prime}+|C_9^{\rm{eff}}|^2
B_2^{\prime} +|C_7^{\rm{eff}}|^2
B_3^{\prime}+2Re[C_9^{\rm{eff}}C_7^{\rm{eff}*}] B_4^{\prime}
\nonumber\\
&& =|C_{10}|^2
B_1^{\prime}+[|C_9|^2+|C_9^{\prime}|^2+2Re[C_9C_9^{\prime *}]]
B_2^{\prime} +|C_7^{\rm{eff}}|^2
B_3^{\prime}+2Re[(C_9+C_9^{\prime})C_7^{\rm{eff}*}]
B_4^{\prime} ~.
\label{eq:BrDecomposition}
\end{eqnarray}
After the integration over $q^2$, Eq.~(\ref{eq:BrDecomposition}) can be rearranged as
\begin{eqnarray}
Br(B\to K\ell^+\ell^-)&=&|C_{10}|^2 B_1+|C_9|^2 B_2 +|C_7^{\rm{eff}}|^2
B_3+2Re[C_9C_7^{\rm{eff}*}] B_4+2Re[C_9] B_5 \nonumber \\
&& + 2Re[C_7^{\rm{eff}}] B_6 + B_7 ~,
\end{eqnarray}
where $B_5$ ($B_6$, $B_7$) contains the integration of $Re[C_9^{\prime}]B_2^{\prime}$
 ($Re[C_9^{\prime}]B_4^{\prime}$, $|C_9^{\prime}|^2B_2^{\prime}$).  Similarly,
 $Br(B\to K^*\ell^+\ell^-)$ is decomposed as
\begin{eqnarray}
Br(B\to K^*\ell^+\ell^-)&=&|C_{10}|^2 B_1^*+|C_9|^2 B_2^*
+|C_7^{\rm{eff}}|^2 B_3^*+2Re[C_9C_7^{\rm{eff}*}] B_4^*+2Re[C_9]
B_5^* \nonumber \\
&& + 2Re[C_7^{\rm{eff}}] B_6^* + B_7^* ~.
\label{eq:Brs}
\end{eqnarray}
The values of $B^{(*)}_j$ with $j=1,2,3, ..., 7$ are, in units of $10^{-8}$ ($10^{-7}$),
\begin{eqnarray}
B_1&=&1.28_{-0.23}^{+0.30} \;,\; B_2=B_1 \;,\;
B_3=4.41_{-0.82}^{+1.44} \;,\; B_4=2.33_{-0.39}^{+0.71} \;,\;\nonumber \\
B_5&=&0.31_{-0.05}^{+0.09}\;,\; B_6=0.58_{-0.10}^{+0.19}\;,\;
B_7=0.18_{-0.03}^{+0.04}\;,\;\nonumber \\
B_1^*&=&0.41_{-0.15}^{+0.20}\;,\; B_2^*=B_1^*\;,\;
B_3^*=12.74_{-4.86}^{+6.35}\;,\; B_4^*=0.84_{-0.45}^{+0.46}\;,\nonumber \\
B_5^*&=&0.09_{-0.03}^{+0.04}\;,\; B_6^*=0.18_{-0.10}^{+0.10}\;,\;
B_7^*=0.04_{-0.02}^{+0.02} ~.
\label{eq:Bis}
\end{eqnarray}
These values will be used to constrain the couplings in the $Z^{\prime}$ model later.  From
Fig.~\ref{Fig:BtoK} and Eqs.~(\ref{eq:HL0}) to (\ref{eq:HR-}), one can find a pole at $q^2=0$
in $dBr(B\to K^*\ell^+\ell^-)/dq^2$.  That is why $B_3^*$ is much larger than the others.

%-------------------------------------------------------------------------
\subsection{The forward-backward asymmetry}
\label{sec:FBS}
%-------------------------------------------------------------------------

The differential forward-backward asymmetry of $\bar B\to \bar K^*\ell^+\ell^-$ is defined by
\begin{eqnarray}
 \frac{d A_{FB}}{dq^2}&=&{\int_0^1 d\cos\theta_1 \frac{d^2\Gamma}{dq^2 d\cos\theta_1}
 -\int_{-1}^0 d\cos\theta_1 \frac{d^2\Gamma}{dq^2 d\cos\theta_1}},
\end{eqnarray}
while the normalized differential forward-backward asymmetry is
defined by
\begin{eqnarray}
  && \frac{d \bar A_{FB}}{dq^2} =
  \frac{\frac{d A_{FB}}{dq^2}}{ \frac{d\Gamma}{dq^2}} \nonumber \\
  && =
  \frac{3}{4} \frac{-|H(L,+)|^2+|H(R,+)|^2+|H(L,-)|^2-|H(R,-)|^2}
  {|H(L,0)|^2+|H(R,0)|^2+|H(L,+)|^2+|H(R,+)|^2+|H(L,-)|^2+|H(R,-)|^2} ~.
  \label{eq:AFB}
\end{eqnarray}
Substituting the expressions in Eqs.~(\ref{eq:HL0}) to (\ref{eq:HR-})
into Eq.~(\ref{eq:AFB}), we get the
explicit expression for $\frac{d \bar A_{FB}}{dq^2}$ as follows:
\begin{eqnarray}
\frac{d \bar A_{FB}}{dq^2}=\frac{3N(q^2)}{4D(q^2)} ~,
\end{eqnarray}
where
\begin{eqnarray}
 N(q^2) &=& |V_{tb}|^2|V_{ts}^*|^2G_F^2\alpha_{\rm{em}}^2\sqrt{\lambda}q^2
 \left\{-Re[C_{10}]C_7^{\rm{eff}}m_b \left[(m_B+m_{K^*})A_1(q^2)T_1(q^2)
     \right. \right. \nonumber \\
 && \left.\left. +(m_B-m_{K^*})T_2(q^2)V(q^2) \right]
 +Re[C_9^{\rm{eff}}C_{10}^*][-q^2V(q^2)A_1(q^2)]\right\} ~,
 \nonumber \\
 D(q^2) &=& 2\pi^2(q^2)^2 \left[ |H(L,0)|^2+|H(R,0)|^2+|H(L,+)|^2+|H(R,+)|^2 \right. \nonumber \\
&& \left. +|H(L,-)|^2+|H(R,-)|^2 \right] ~.
 \label{eq:AFBexpression}
\end{eqnarray}
In the above expression, terms suppressed by $m_s$ are dropped for simplicity.  As can be
explicitly checked, the pole in the dilepton spectrum at $q^2=0$ disappears in the denominator.

%--------------------------------------
\begin{figure}[htb]
  \includegraphics[width=8cm]{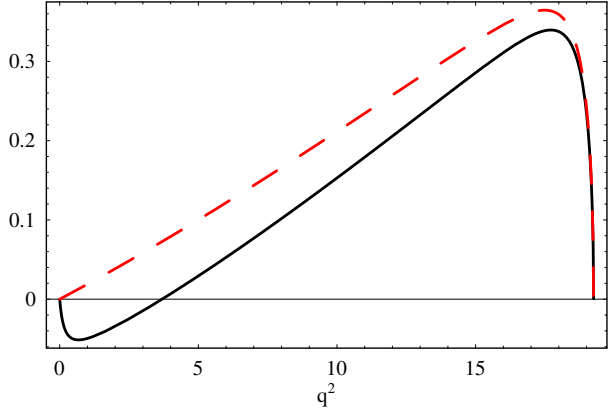}
  \vspace{-1.5cm}
  \caption{The forward-backward asymmetry for $B\to K^*\ell^+\ell^-$ with form factors given by
  the PQCD approach.  The black solid curve is given with the SM $C_7$ and the red dashed curve
  is given with $C_7=0$.}
  \label{fig:FBSM}
\end{figure}
%--------------------------------------

According to Eq.~(\ref{eq:AFBexpression}), the numerator of $dA_{FB}/dq^2$ is zero at $q^2=0$
because of the common factor $q^2$, while the denominator has a non-zero value because its
common factor $(q^2)^2$ cancels with the $(q^2)^2$ factor arising from Eqs.~(\ref{eq:HL+}),
(\ref{eq:HL-}), (\ref{eq:HR+}) and (\ref{eq:HR-}).  Thus $dA_{FB}/dq^2=0$ at $q^2=0$. In the
 SM, $C_7^{\rm{eff}}<0$, $C_9^{\rm{eff}}>0$, and $C_{10}<0$; thus the first term in the curly
  bracket of $N(q^2)$ is negative and the second term is positive.  In the regime where $q^2$
  is near zero, the first term gives the dominant contribution since the second term is
   suppressed by the small $q^2$.  Therefore, the sign of $dA_{FB}/dq^2$ is determined by the
   first term and gives a negative value.  As $q^2$ increases, the second term becomes dominant.
    There exists a point where $dA_{FB}/dq^2$ becomes zero, the so-called forward-backward
    asymmetry zero.  The position of the zero is determined by $C_7^{\rm eff}$ and $C_9^{\rm eff}$,
     for the form-factor dependence drops at the leading order \cite{SM-NP}.  As $q^2$ becomes
     even larger, the effect of the overall factor $\sqrt{\lambda}$ becomes crucial.
      Eq.~(\ref{eq:lambda}) tells us that $\lambda=0$ at the largest recoil where
      $q^2=(m_B-m_{K^*})^2$.  Therefore, $dA_{FB}/dq^2$ falls back to zero at the end
       of the kinematic regime.
 All these behaviors of $dA_{FB}/dq^2$ can be observed in Fig.~\ref{fig:FBSM}.  The red
 dashed curve is drawn with the contribution of only the second term in the curly bracket
 of $N(q^2)$.  It shows the importance of $C_7^{\rm{eff}}$ in the low $q^2$ regime.

However, the latest Belle data \cite{FBSexperiment} do not show an
obvious zero for $dA_{FB}/dq^2$, and the values at all $q^2$ are
consistently higher than the SM expectation.  A common solution is
to flip the sign of $C_7^{\rm eff}$ as it is still consistent with
the constraint from $B \to X_s \gamma$ data.  In the next section,
we offer an alternative solution in the family non-universal $Z'$
model.

%=====================================================================================
\section{Constraints on the couplings in $Z^{\prime}$ physics}
\label{sec:Zprime}
%=====================================================================================

%-------------------------------------------------------------
\subsection{$b\to s\ell^+\ell^-$ in the $Z^{\prime}$ FCNC model}
%-------------------------------------------------------------

In the appropriate gauge basis, the $U(1)^{\prime}$ currents
are
\begin{eqnarray}
J_{Z^{\prime}}^{\mu}=g^{\prime}\sum_i \bar\psi_i
\gamma^{\mu}[\epsilon_i^{\psi_L}P_L+\epsilon_i^{\psi_R}P_R]\psi_i,
\label{eq:JZprime}
\end{eqnarray}
where $i$ is the family index and $\psi$ labels the fermions (up- or down-type quarks,
 or charged or neutral leptons), and $P_{L,R}=(1\mp\gamma_5)/2$.
 According to some string construction or GUT models such as $E_6$, it is possible to have
  family non-universal $Z^{\prime}$ couplings.  That is, even though $\epsilon_i^{L,R}$ are diagonal,
   the couplings are not family universal. After rotating to the physical basis, FCNCs generally
   appear at tree level in both LH and RH sectors.  Explicitly,
\begin{eqnarray}
B^{\psi_L}=V_{\psi_L}\epsilon^{\psi_L}V_{\psi_L}^{\dagger},\;\;\;\;\;
B^{\psi_R}=V_{\psi_R}\epsilon^{\psi_R}V_{\psi_R}^{\dagger}.
\end{eqnarray}
Moreover, these couplings may contain CP-violating phases beyond that of the SM.

In particular, $Z^{\prime}\bar bs$ couplings can be generated:
\begin{eqnarray}
{\cal L}_{\rm{FCNC}}^{Z^{\prime}}=-g^{\prime}(B_{sb}^L\bar
s_L\gamma_{\mu}b_L + B_{sb}^R\bar
s_R\gamma_{\mu}b_R)Z^{\prime\mu} + {\rm h.c.} ~.
\label{eq:HamiltonZ}
\end{eqnarray}
%These couplings lead to a new physics contribution to $B_s^0-\bar
%B_s^0$ mixing at tree level. Let's define
%\begin{eqnarray}
%\rho_{ff^{\prime}}^{L,R}\equiv
%\left|\frac{g^{\prime}M_Z}{gM_{Z^{\prime}}}B_{ff^{\prime}}^{L,R}\right|.
%\end{eqnarray}
The couplings in Eq.~(\ref{eq:HamiltonZ}) lead to extra
contributions to the $b\to s\ell^+\ell^-$ decay at tree level,
mediated by a virtual $Z^{\prime}$ boson.  The amplitude is given by
\begin{eqnarray}
\frac{g^{\prime 2}}{M_{Z^{\prime}}^2}
\left(B_{sb}^L\bar s_L\gamma_{\mu}b_L + B_{sb}^R\bar s_R\gamma_{\mu}b_R\right)
\left(B_{\ell\ell}^L\bar \ell_L\gamma^{\mu}l_L + B_{\ell\ell}^R\bar \ell_R\gamma^{\mu} \ell_R\right)
 ~.
\end{eqnarray}
There are thus four types of operators, $O_{LL}$, $O_{LR}$,
$O_{RL}$, and $O_{RR}$. The above amplitude can be derived from an effective
Hamiltonian
\begin{eqnarray}
 {\cal H}_{\rm{eff}}^{Z^{\prime}}=
\frac{8G_F}{\sqrt{2}} (\rho_{sb}^L\bar s_L\gamma_{\mu}b_L +
\rho_{sb}^R\bar s_R\gamma_{\mu}b_R) (\rho_{ll}^L\bar
\ell_L\gamma^{\mu}\ell_L +\rho_{ll}^R\bar \ell_R\gamma^{\mu}\ell_R)
~, \label{eq:HamiltonZprime}
\end{eqnarray}
where
\begin{eqnarray}
\rho_{ff'}^{L,R} \equiv
\frac{g'M_Z}{gM_{Z'}} B_{ff'}^{L,R}
\end{eqnarray}
and $g$ is the coupling associated with the $SU(2)_L$ group in the SM.
%CC added following sentence
Throughout this analysis, we ignore the renormalization group
running effects due to these new contributions because they are
expected to be small.

%-------------------------------------------------------------
\subsection{Constraints from the $B\to K^{(*)}\ell^+\ell^-$ decays}
%-------------------------------------------------------------

%%%%%%%%%%%%%%%%%%%%%%%%%%%%%%%%%%%%%%%%%%%%%%%%%%%%%%%%%%%%%%%%%%%%%%%%%%%
 \begin{table}
 \caption{Values of Wilson coefficients $C_i(m_b)$ in the leading
logarithmic approximation, with $m_W=80.4\mbox{GeV}$, $\mu=m_{b,\rm
pole}$~\cite{Buchalla:1995vs}.}
 \label{tab:wilsons}
 \begin{center}
 \begin{tabular}{c c c c c c c c c}
 \hline\hline
 \ \ \ $C_1$ &$C_2$ &$C_3$ &$C_4$ &$C_5$ &$C_6$ &$C_7^{\rm{eff}}$ &$C_9$ &$C_{10}$       \\
\hline
 \ \ \ $1.107$   &$-0.248$   &$-0.011$    &$-0.026$    &$-0.007$    &$-0.031$    &$-0.313$    &$4.344$    &$-4.669$    \\
 \hline\hline
 \end{tabular}
 \end{center}
 \end{table}
%%%%%%%%%%%%%%%%%%%%%%%%%%%%%%%%%%%%%%%%%%%%%%%%%%%%%%%%%%%%%

For the purpose of illustration and to avoid too many free parameters, we assume that the FCNC
couplings of the $Z^{\prime}$ and quarks only occur in the left-handed (LH) sector.  Therefore,
$\rho_{sb}^R=0$, and the effects of the $Z^{\prime}$ FCNC currents simply modify the Wilson
coefficients $C_9$ and $C_{10}$ in Eq.~(\ref{eq:Hamiltonian}).  We denote these two modified
Wilson coefficients by $C_9^{\rm{eff},Z^{\prime}}$ and $C_{10}^{Z^{\prime}}$, respectively.
 More explicitly,
\begin{eqnarray}
 Re[C_9^{\rm{eff},Z^{\prime}}]&=&Re[C_9^{\rm{eff}}]-{4\pi Re[\rho_{sb}^L](\rho_{ll}^L+\rho_{ll}^R)
 \over V_{tb}V^*_{ts}\alpha_{em}},\nonumber\\
 Im[C_9^{\rm{eff},Z^{\prime}}]&=&Im[C_9^{\rm{eff}}]-{4\pi Im[\rho_{sb}^L](\rho_{ll}^L+\rho_{ll}^R)
  \over V_{tb}V^*_{ts}\alpha_{em}},\nonumber\\
 Re[C_{10}^{Z^{\prime}}]&=&C_{10}-{4\pi Re[\rho_{sb}^L](\rho_{ll}^R-\rho_{ll}^L) \over V_{tb}V^*_{ts}
 \alpha_{em}},\nonumber\\
 Im[C_{10}^{Z^{\prime}}]&=&-{4\pi Im[\rho_{sb}^L](\rho_{ll}^R-\rho_{ll}^L) \over
 V_{tb}V^*_{ts}\alpha_{em}}.\label{eq:wilsonZprime}
\end{eqnarray}
For simplicity, we further assume that $\rho_{sb}^L$ is real.  Then the imaginary part of
 $C_9^{\rm{eff}}$ will not be affected by the $Z^{\prime}$ model, and $C_{10}^{Z^{\prime}}$
 is still a real number.

First, we consider the constraint from the spectrum of $d\bar A_{FB}/dq^2$.  In order to fit
 the experimental data, a sign flip is needed for $d\bar A_{FB}/dq^2$ near the $q^2=0$ regime.
  People usually consider
the flipped-sign solution with $C_7 = - C_7^{\rm SM}$, because it is
still allowed by the $B \to X_s \gamma$ data.  However, an
alternative solution is to flip the signs of
 $C_9^{\rm{eff}}$ and $C_{10}$ instead, as is possible in our model.  Below
 Eq.~(\ref{eq:AFBexpression}), it is noted that in this regime the term proportional
  to $Re[C_{10}]C_7^{\rm{eff}}$ dominates.  Therefore, one can flip the sign of $C_{10}$:
\begin{equation}
Re[C_{10}^{Z^{\prime}}]>0 ~.
\label{eq:condition1}
\end{equation}
Moreover, in order to keep the second term in the curly bracket of $N(q^2)$ to have the correct
 behavior, we also need to flip the sign of $Re[C_9^{\rm{eff}}]$. Thus, we require
\begin{equation}
Re[C_9^{\rm{eff},Z^{\prime}}]<0 ~.
\label{eq:condition2}
\end{equation}
Eqs.~(\ref{eq:condition1}) and (\ref{eq:condition2}) are the constraints from the
 $d\bar A_{FB}/dq^2$ spectrum obtained by the Belle Collaboration
(see Fig.~1 in Ref.~\cite{FBSexperiment}).

%{\bf Are there any other modes that are sensitive to the signs of $C_{9,10}$?}

Next, we consider the constraints from the branching ratios of $B \to K^{(*)} \ell^+ \ell^-$ decays.
  These constraints are obtained in the following way.
 After including the contributions of $Z^{\prime}$, the upper (lower) bound of the theoretical
  predictions should be greater (smaller) than the experimental lower (upper) bound at the
  $2\sigma$ level.  When we deal with the experimental data, we add the statistical and systematic
   errors in quadrature.  With Eqs.~(\ref{eq:Brs}) and (\ref{eq:Bis}), we have the following
   branching-ratio constraints:
\begin{eqnarray}
B^{(*)}_{1u}(|C_{10}^{Z^{\prime}}|^2+|C_9^{Z^{\prime}}|^2)+B^{(*)}_{3u}|C_7^{\rm{eff}}|^2
+B^{(*)}_{4u}Re[C_9^{Z^{\prime}}C_7^{\rm{eff}*}]
+B^{(*)}_{5u}Re[C_9^{Z^{\prime}}] \nonumber \\\
\qquad\qquad + B^{(*)}_{6u}Re[C_7^{\rm{eff}}] + B^{(*)}_{7u}
> Br^{(*)}_{\rm{exp}}-2\sigma^{(*)}_l\;,\label{eq:condition3}\\
B^{(*)}_{1l}(|C_{10}^{Z^{\prime}}|^2+|C_9^{Z^{\prime}}|^2)+B^{(*)}_{3l}|C_7^{\rm{eff}}|^2
+B^{(*)}_{4l}Re[C_9^{Z^{\prime}}C_7^{\rm{eff}*}]
+B^{(*)}_{5l}Re[C_9^{Z^{\prime}}] \nonumber \\
\qquad\qquad +B^{(*)}_{6l}Re[C_7^{\rm{eff}}]+B^{(*)}_{7l}
< Br^{(*)}_{\rm{exp}}+2\sigma^{(*)}_u\;,\label{eq:condition4}
\end{eqnarray}
%\begin{eqnarray}
%B\to K\ell^+\ell^-\bigg\{
%\begin{array}{cc}
%1.58(|C_{10}^{Z^{\prime}}|^2+|C_9^{Z^{\prime}}|^2)+5.85|C_7^{\rm{eff}}|^2+6.08Re[C_9^{Z^{\prime}}C_7^{\rm{eff}*}]
%+0.80Re[C_9] + 1.54Re[C_7^{\rm{eff}}] + 0.22 >38.00\;,\\
%1.05(|C_{10}^{Z^{\prime}}|^2+|C_9^{Z^{\prime}}|^2)+3.59|C_7^{\rm{eff}}|^2+3.88Re[C_9^{Z^{\prime}}C_7^{\rm{eff}*}]
%+0.52Re[C_9] + 0.96Re[C_7^{\rm{eff}}] + 0.15 <59.67\;,
%\end{array}\label{eq:condition3}\\
%B\to K^*\ell^+\ell^-\bigg\{
%\begin{array}{cc}
%0.61(|C_{10}^{Z^{\prime}}|^2+|C_9^{Z^{\prime}}|^2)+19.09|C_7^{\rm{eff}}|^2+2.60Re[C_9^{Z^{\prime}}C_7^{\rm{eff}*}]
%+0.26 Re[C_9] + 0.56 Re[C_7^{\rm{eff}}] + 0.06 >8.01\;,\\
%0.26(|C_{10}^{Z^{\prime}}|^2+|C_9^{Z^{\prime}}|^2)+7.88|C_7^{\rm{eff}}|^2+0.78Re[C_9^{Z^{\prime}}C_7^{\rm{eff}*}]
%+0.12 Re[C_9] + 0.16 Re[C_7^{\rm{eff}}] + 0.02 <13.54\;,
%\end{array}\label{eq:condition4}
%\end{eqnarray}
where quantities with a star in the superscript are for the $B\to
K^*\ell^+\ell^-$ decay, and the letters ``$u$'' and ``$l$'' in the
subscript represent the 1-$\sigma$ upper and lower bounds of the
corresponding quantity $B_i^{(*)}$, respectively, and
$Br^{(*)}_{\rm{exp}}$ denote the central values of the $B\to K^{(*)}
\ell^+\ell^-$ branching ratios.

\begin{figure}
  \includegraphics[width=8cm]{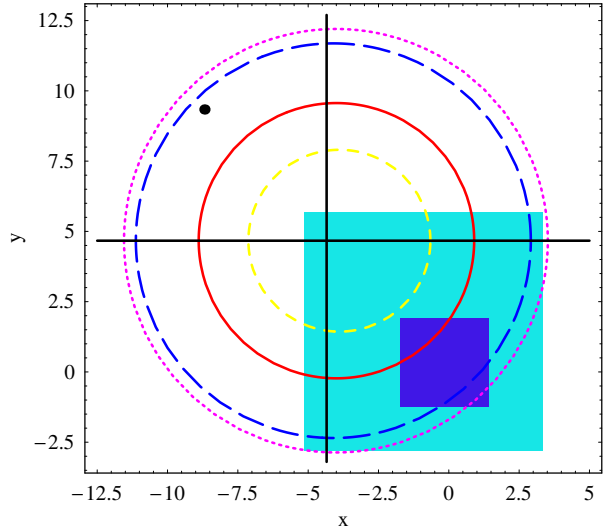}
  \caption{The constraints from branching ratios of $B \to K^{(*)} \ell^+ \ell^-$ decays.
  The areas outside the red solid and yellow short dashed circles are determined by
  Eqs.~(\ref{eq:con3}) and (\ref{eq:con5}), respectively.  The areas inside the pink and
  blue circles are determined by Eqs.~(\ref{eq:con4}) and (\ref{eq:con6}), respectively.
   The areas to the left of the line $x=-C_9$ and above the line $y=-C_{10}$ are determined
   by Eqs.~(\ref{eq:con1}) and (\ref{eq:con2}), respectively. The black dot is where both
   $C_9$ and $C_{10}$ flip  signs from their SM values. The two rectangles, corresponding
   to S1 (the large rectangle) and S2 (the small rectangle) in Case III, are the constraints
   given by Ref.~\cite{ZprPheno5}. One can see that their constraints are consistent with
   our constraints from the branching ratios. However, their constraints are not enough to
   change the signs of $C_9$ and $C_{10}$.}
 \label{fig:constraints}
\end{figure}

Moreover, $C_9^{Z^{\prime}}=C_9+x$ with $Y_{\rm{pert}}(q^2)$ excluded, and $C_{10}^{Z'} = C_{10}
 + y$, where
\begin{eqnarray}
x&=&-{4\pi Re[\rho_{sb}^L](\rho_{ll}^L+\rho_{ll}^R) \over V_{tb}V^*_{ts}\alpha_{em}}\;,\label{eq:x}\\
y&=&-{4\pi Re[\rho_{sb}^L](\rho_{ll}^R-\rho_{ll}^L) \over
V_{tb}V^*_{ts}\alpha_{em}} ~. \label{eq:y}
\end{eqnarray}
%we have
%\begin{eqnarray}
%Re[C_9^{\rm{eff},Z^{\prime}}]=Re[C_9^{\rm{eff}}]+x\;,\;Re[C_{10}^{\rm{eff},Z^{\prime}}]=Re[C_{10}^{\rm{eff}}]+y.\label{eq:Wils}
%\end{eqnarray}
Then Eqs. (\ref{eq:condition3}) and (\ref{eq:condition4}) can be
rearranged as
\begin{eqnarray}
B_{1u}^{(*)}(x+T_u^{(*)})^2+B_{1u}^{(*)}(y+C_{10})^2+C_u^{(*)}&>&Br^{(*)}_{\rm{exp}}-2\sigma^{(*)}_l ~,\label{eq:cona1}\\
B_{1l}^{(*)}(x+T_l^{(*)})^2+B_{1l}^{(*)}(y+C_{10})^2+C_l^{(*)}&<&Br^{(*)}_{\rm{exp}}+2\sigma^{(*)}_u ~,\label{eq:cona2}
\end{eqnarray}
where
\begin{eqnarray}
T_{u/l}^{(*)}&=&\frac{2B_{1u/l}^{(*)}C_9+B_{4u/l}^{(*)}C_7^{\rm{eff}}+B_{5u/l}^{(*)}}{2B_{1u/l}^{(*)}},\nonumber\\
C_{u/l}^{(*)}&=&B_{1u/l}^{(*)}C_9^2+B_{7u/l}^{(*)}(C_7^{\rm{eff}})^2+B_{4u/l}^{(*)}C_7^{\rm{eff}}C_9
+B_{5u/l}^{(*)}C_9+B_{6u/l}^{(*)}C_7 \nonumber \\
&& +B_{7u/l}^{(*)} -
B_{1u/l}^{(*)} \big(T_{u/l}^{(*)}\big)^2.
\end{eqnarray}
Substituting all the numerical values in Eqs. (\ref{eq:condition1}),
(\ref{eq:condition2}), (\ref{eq:cona1}), and (\ref{eq:cona2}), we
have
\begin{eqnarray}
x&<&-4.344 ~,\label{eq:con1}\\
y&>&4.669 ~,\label{eq:con2}\\
1.58(x+3.99)^2+1.58(y-4.669)^2-37.88&>&0 ~,\label{eq:con3}\\
1.05(x+4.01)^2+1.05(y-4.669)^2-59.58&<&0 ~,\label{eq:con4}\\
0.61(x+3.89)^2+0.61(y-4.669)^2-6.38&>&0 ~,\label{eq:con5}\\
0.26(x+4.11)^2+0.26(y-4.669)^2-12.81&<&0 ~.\label{eq:con6}
\end{eqnarray}
Eqs.~(\ref{eq:con1})-(\ref{eq:con6}) give the constraints on $x$ and
$y$, which are shown in Fig.~\ref{fig:constraints}. The common area
of the above six conditions is outside the red solid circle and
inside the blue long dashed circle, to the left of the solid
vertical line $x=-C_9$ and above the solid horizontal line
$y=-C_{10}$.  This area gives

\begin{eqnarray}
&
-\sqrt{(Br^*_{\rm{exp}}+2\sigma^*_u-C^*_l)/B_{1l}^*}-T_l^* \lesssim
x \lesssim -C_9 ~,
& \nonumber \\
& -C_{10}\lesssim y \lesssim
\sqrt{(Br^*_{\rm{exp}}+2\sigma^*_u-C^*_l)/B_{1l}^*}-C_{10} ~. &
\label{eq:constraintxy}
\end{eqnarray}
With Eqs. (\ref{eq:x}) and (\ref{eq:y}), we have
\begin{eqnarray}
&& \left[\sqrt{(Br^*_{\rm{exp}}+2\sigma^*_u-C^*_l)/B_{1l}^*}-C_{10}-C_9\right]{\cal
K} \lesssim Re[\rho_{sb}^L]\rho_{ll}^R \nonumber \\
&& \qquad\qquad \lesssim
\left[-\sqrt{(Br^*_{\rm{exp}}+2\sigma^*_u-C^*_l)/B_{1l}^*}-T_l^*
-C_{10}\right]{\cal K} \;,\nonumber\\
&& \left[C_{10}-C_9\right]{\cal K} \lesssim Re[\rho_{sb}^L]\rho_{ll}^L \nonumber \\
&& \qquad\qquad \lesssim
\left[-2\sqrt{(Br^*_{\rm{exp}}+2\sigma^*_u-C^*_l)/B_{1l}^*}-T_l^*+C_{10}\right]
{\cal K} \;,\label{eq:LRcoupling}
\end{eqnarray}
with ${\cal K}=(V_{tb}V^*_{ts}\alpha_{\rm{em}})/(4\pi)$. In the
quark sector, the couplings in Eq.~(\ref{eq:HamiltonZ}) also lead to
a NP contribution to $B_s^0-\bar B_s^0$ mixing at tree level.  In
Refs.~\cite{rhosbL,rhosbL2}, it is assumed that only the LH sector
of quarks has family non-universal $U(1)^{\prime}$ couplings, as in the current analysis.  Thus,
only the LH interaction in Eq.~(\ref{eq:HamiltonZ}) contributes to
the $B_s^0-\bar B_s^0$ mixing.  They find that one can reproduce the
measured value of $\Delta M_s$ if
\begin{eqnarray}
\rho_{sb}^L\lesssim 10^{-3} ~.
\end{eqnarray}
As a rough estimate, here we take $\rho_{sb}^L=10^{-3}$.  Together
with Eqs.~(\ref{eq:x}), (\ref{eq:y}), and
\begin{eqnarray}
V_{tb}=0.999176\;,\;V_{ts}=-0.03972\;,\;\alpha_{em}=1/137 ~,
\end{eqnarray}
we obtain
\begin{eqnarray}
-0.27&\lesssim&\rho_{ll}^L\lesssim -0.11 ~,\\
-0.08&\lesssim&\rho_{ll}^R\lesssim 0.09 ~.
\end{eqnarray}
We should emphasize that these parameter ranges are obtained with
some assumptions and the current data.  In particular, we have used
a particular value of $\rho_{sb}^L$ for our illustration.   Once new
experimental data or theoretical inputs are available, these
constraints can be easily updated with our formulas. In
Fig.\ref{fig:constraints} we also give the constraints from
Ref.~\cite{ZprPheno5}. In their paper the authors gain the
constraints by making the experimental and theoretical values of
$B\to X_s\l^+l^-$ agree with each other in $1\sigma$. However, we
get our constraints in $2\sigma$. For a comparison, in
Fig.\ref{fig:constraints} we simply extrapolate their results to
$2\sigma$. One can find that if we drop the constraint conditions of
flipping the signs of $C_9^{\rm{eff}}$ and $C_{10}$, we agree with
each other. However, if the $A_{FB}$ is expected to behave as how we
get constraints (\ref{eq:condition1}) and (\ref{eq:condition2}), the
constraints in Ref.~\cite{ZprPheno5} are too tight to satisfy the
conditions.

%------------------------------------------------------------------
\begin{figure*}[htb]
 \includegraphics[scale=0.56]{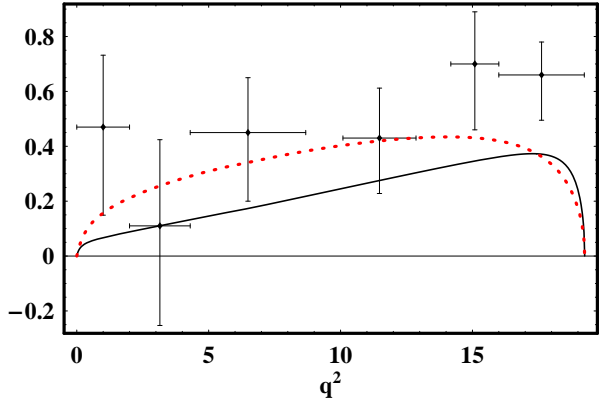}
 \vspace{-1.5cm}
 \caption{Forward-backward asymmetry in QCDSR (red dotted line)
  and PQCD (black solid line) with $C_9$ and $C_{10}$ flipping
  their signs (the black dot in Fig.~\ref{fig:constraints}).
  The points with error bars are the experimental results
  from the Belle Collaboration \cite{FBSexperiment}.}
 \label{fig:FBZprime}
\end{figure*}
%------------------------------------------------------------------

In Fig.~\ref{fig:FBZprime}, we use the black dot from
Fig.~\ref{fig:constraints}, where both $C_9^{\rm{eff}}(q^2)$ and
$C_{10}$ flip  signs from their SM values, to predict the
$dA_{FB}/dq^2$ spectrum in our model. Since $C_9^{\rm{eff}}(q^2)$ is
$q^2$-dependent, the plot in Fig.~\ref{fig:constraints} is plotted
with $C_9$ and $C_{10}$ flipping their signs. The points that flip
the signs of $C_9^{\rm{eff}}(q^2)$ and $C_{10}$ should be very close
to this point. It is interesting to note that the red dotted curve
in Fig.~\ref{fig:FBZprime} is identical to the usual flipped-sign
solution.  This is not surprising because flipping the signs of both
$C_9^{\rm{eff}}(q^2)$ and $C_{10}$ simultaneously is equivalent to
flipping the sign of $C_7^{\rm{eff}}$, which can be seen from
Eq.~(\ref{eq:AFBexpression}). This indicates that by considering
only the branching ratios and forward-backward asymmetry of the $B
\to K^{(*)} \ell^+ \ell^-$ decays, it is insufficient to determine
which operators are significantly modified by the NP.

Now a comment on the form factors is in order.  Because of the nonperturbative effects,
we cannot get good results for the form factors when $q^2$ is large.  In either PQCD or
 light cone sum rules, the form factors are obtained in a region where $q^2$ is small
 and then extrapolated to the entire kinematical region through fitting.  As a result,
 it is a question whether the form factors can be described well by the parametrization
  formula in the large $q^2$ region.  In fact, the accuracy of the parametrization formula
   becomes worse as $q^2$ increases. Therefore, we do not think the theoretical predictions
    at large $q^2$ are reliable enough.  This may explain why the experimental values
    are still a little larger than the theoretical predictions in the large $q^2$ regime,
     as shown in Fig.~\ref{fig:FBZprime}.

A closely related decay mode to the current analysis is the $B_s\to \mu^+\mu^-$ decay.
 This mode has been searched for with great interest at Tevatron.  The upper bounds on
 the branching ratio at $95\%$ confidence level are given by its two experimental groups as
\begin{eqnarray}
Br(B_s\to \mu^+\mu^-)&<& 5.8\times 10^{-8}\;\;\;\;\;\mbox{(CDF) \cite{Aaltonen:2007kv}} ~,\nonumber\\
Br(B_s\to \mu^+\mu^-)&<& 1.2\times
10^{-7}\;\;\;\;\;\mbox{(D{\O}) \cite{Abazov:2007iy}} ~.
\end{eqnarray}
The branching ratio of $B_s\to \mu^+\mu^-$ is affected in our model.  With the inclusion of the
 $Z'$ contribution, the branching ratio is given
by\cite{Buchalla:1995vs}
\begin{eqnarray}
Br(B_s\to\mu^+\mu^-) &=& \tau_{B_s} \frac{G_F^2} {4\pi} f_{B_s}^2
m_{\mu}^2 m_{B_s} \sqrt{1-\frac{4m_{\mu}^2}{m_{B_s}^2}}
\left|V^*_{tb}V_{ts}\right|^2 \nonumber \\
&& \times \left|\frac{\alpha}{2\pi\sin^2\theta_W}Y\left(\frac{m_t^2}{m_W^2}\right)
+ 2 \frac{\rho^L_{bs}(\rho_{\mu\mu}^L-\rho_{\mu\mu}^R)}
{V^*_{tb}V_{ts}} \right|^2,
\end{eqnarray}
where all the functions and symbols are defined in
Ref.~\cite{Buchalla:1995vs}. With the constraints in
Eq.~(\ref{eq:constraintxy}), we find that the upper bound for this
branching ratio is
\begin{eqnarray}
Br(B_s\to \mu^+\mu^-) \lesssim 7.9\times 10^{-9}.
\end{eqnarray}
Note that the upper bound of the range is still smaller than the
current upper bound given by CDF Collaboration.

%=====================================================
\section{summary \label{sec:summary}}
%=====================================================

We have considered the contributions of family non-universal $Z^{\prime}$ models with
flavor-changing neutral currents ($Z'$ FCNC) at tree level in $B\to K^{(*)}\ell^+\ell^-$
decays.  By requiring that the theoretically predicted branching ratios agree with the
current experimental data within two $\sigma$'s, we obtain the constraints on the couplings
 in the $Z^{\prime}$ FCNC model.  We find that within the allowed parameter space, our model
 has the potential to explain the forward-backward asymmetry of the $B\to K^*\ell^+\ell^-$
 decay, as better determined by the Belle Collaboration recently.  Moreover, our
 Z' model contributions flip the signs of
  $C_9^{\rm eff}$ and $C_{10}$, which
 differs from the usual   new physics contributions that flip the sign of
   $C_7^{\rm eff}$.  Using the constraints, we also compute the branching ratio of the
   $B_s\to\mu^+\mu^-$ decay. The upper bound of our prediction is near the upper
   bound given by CDF Collaboration.

%=====================================================
\section{acknowledgments}
%=====================================================

C.-W.~C. would like to thank the hospitality of IHEP, Beijing, where
this project is initiated, during his visit.

%***********************************************************************************
\begin{appendix}
%=================================================================================
 \section{functions for the leptonic and hadronic part}
 \label{appendix:LH}
%=================================================================================

\begin{eqnarray}
 {L}(L,0)&=& 2\sqrt {q^2}\sin\theta_1,\;\;\;\\
 {L}(L,+)&=& -2\sqrt 2\sqrt {q^2}\sin^2\frac{\theta_1}{2}
 e^{i\phi},\;\;\;\\
 {L}(L,-)&=& -2\sqrt 2\sqrt {q^2}\cos^2\frac{\theta_1}{2}
 e^{-i\phi},\\
 {L}(R,0)&=& -2\sqrt {q^2}\sin\theta_1,\;\;\;\\
 {L}(R,+)&=& -2\sqrt 2\sqrt {q^2}\cos^2\frac{\theta_1}{2}
 e^{i\phi},\;\;\;\\
 {L}(R,-)&=& -2\sqrt 2\sqrt {q^2}\sin^2\frac{\theta_1}{2}
 e^{-i\phi}.
\end{eqnarray}

\begin{eqnarray}
 {H}(L,0)&=&\frac{iG_F V_{tb} V^*_{ts} \alpha_{em}}{8\sqrt{2} \pi m_{K^*} \sqrt{q^2}}
 \bigg\{2(C_{7L}-C_{7R})m_b\left[\frac{\lambda T_3(q^2)}{m_B^2-m_{K^*}^2}
 -\left(3m_{K^*}^2+m_B^2-q^2\right)T_2(q^2)\right]\nonumber\\
 &&+(C_9^{\rm eff}-C_{10})\left[(m_B+m_{K^*})(m_{K^*}^2-m_B^2+q^2)A_1(q^2)
 +\frac{\lambda A_2(q^2)}{(m_B+m_{K^*})}\right]\bigg\},\;\;\;\label{eq:HL0}
\end{eqnarray}
\begin{eqnarray}
 {H}(L,+)&=&\frac{iG_F V_{tb} V^*_{ts}\alpha_{em}}{4\sqrt{2}\pi q^2}
 \bigg\{2(C_{7L}+C_{7R})m_b\sqrt{\lambda} T_1(q^2) - 2(C_{7L}-C_{7R}) m_b (m_B^2-m_{K^*}^2) T_2(q^2) \nonumber\\
 &&+(C_9^{\rm eff}-C_{10})q^2\bigg[\frac{\sqrt{\lambda} V(q^2)}{(m_B+m_{K^*})}-(m_B+m_{K^*})A_1(q^2)\bigg]\bigg\},\;\;\;\label{eq:HL+}
\end{eqnarray}
\begin{eqnarray}
 {H}(L,-)&=&\frac{iG_F V_{tb} V^*_{ts}\alpha_{em}}{4\sqrt{2}\pi q^2}
 \bigg\{-2(C_{7L}+C_{7R})m_b\sqrt{\lambda} T_1(q^2) - 2(C_{7L}-C_{7R}) m_b (m_B^2-m_{K^*}^2) T_2(q^2) \nonumber\\
 &&+(C_9^{\rm eff}-C_{10})q^2\bigg[-\frac{\sqrt{\lambda} V(q^2)}{(m_B+m_{K^*})}-(m_B+m_{K^*})A_1(q^2)\bigg]\bigg\},\;\;\;\label{eq:HL-}
\end{eqnarray}
\begin{eqnarray}
 {H}(R,0)&=&\frac{iG_F V_{tb} V^*_{ts} \alpha_{em}}{8\sqrt{2} \pi m_{K^*} \sqrt{q^2}}
 \bigg\{2(C_{7L}-C_{7R})m_b\left[\frac{\lambda T_3(q^2)}{m_B^2-m_{K^*}^2}
 -\left(3m_{K^*}^2+m_B^2-q^2\right)T_2(q^2)\right]\nonumber\\
 &&+(C_9^{\rm eff}+C_{10})\left[(m_B+m_{K^*})(m_{K^*}^2-m_B^2+q^2)A_1(q^2)
 +\frac{\lambda A_2(q^2)}{(m_B+m_{K^*})}\right]\bigg\},\;\;\;\label{eq:HR0}
\end{eqnarray}
\begin{eqnarray}
 {H}(R,+)&=&\frac{iG_F V_{tb} V^*_{ts}\alpha_{em}}{4\sqrt{2}\pi q^2}
 \bigg\{2(C_{7L}+C_{7R})m_b\sqrt{\lambda} T_1(q^2) - 2(C_{7L}-C_{7R}) m_b (m_B^2-m_{K^*}^2) T_2(q^2) \nonumber\\
 &&+(C_9^{\rm eff}+C_{10})q^2\bigg[\frac{\sqrt{\lambda} V(q^2)}{(m_B+m_{K^*})}-(m_B+m_{K^*})A_1(q^2)\bigg]\bigg\},\;\;\;\label{eq:HR+}
\end{eqnarray}
\begin{eqnarray}
 {H}(R,-)&=&\frac{iG_F V_{tb} V^*_{ts}\alpha_{em}}{4\sqrt{2}\pi q^2}
 \bigg\{-2(C_{7L}+C_{7R})m_b\sqrt{\lambda} T_1(q^2) - 2(C_{7L}-C_{7R}) m_b (m_B^2-m_{K^*}^2) T_2(q^2) \nonumber\\
 &&+(C_9^{\rm eff}+C_{10})q^2\bigg[-\frac{\sqrt{\lambda} V(q^2)}{(m_B+m_{K^*})}-(m_B+m_{K^*})A_1(q^2)\bigg]\bigg\},\;\;\;\label{eq:HR-}
\end{eqnarray}

\end{appendix}

%%%%%%%%%%%%%%%%%%%%%%%%%%%%%%

\end{CJK*}

\begin{thebibliography}{11}
%%%%%%%%%%%%%%%%%%%%%%%%%%%%%%

\bibitem{SMbasis}
%\bibitem{Jaus:1989av}
  Jaus W and Wyler D,
  %``The Rare Decays of B $\to$ K Lepton anti-Lepton and B $\to$ K* Lepton
  %anti-Lepton,''
  Phys.\ Rev.\  D, 1990, {\bf 41}, 3405;
  %%CITATION = PHRVA,D41,3405;%%
%\bibitem{Colangelo:1995jv}
 Colangelo P, De Fazio F, Santorelli P and Scrimieri E,
  %``QCD sum rule analysis of the decays $B \to K \ell^{+} \ell^{-}$ and $B \to
  %K^{*} \ell^{+} \ell^{-}$,''
  Phys.\ Rev.\  D, 1996, {\bf 53}, 3672
  [Erratum-ibid.\  D, 1998, {\bf 57}, 3186 ]
  [arXiv:hep-ph/9510403];
  %%CITATION = PHRVA,D53,3672;%%
%\bibitem{Aliev:1996hb}
  Aliev T M, Ozpineci A and Savci M,
  %``Rare B --> K* l+ l- decay in light cone QCD,''
  Phys.\ Rev.\  D, 1997, {\bf 56}, 4260
  [arXiv:hep-ph/9612480];
  %%CITATION = PHRVA,D56,4260;%%
%\bibitem{Melikhov:1998ws}
  Melikhov D, Nikitin N and  Simula S,
  %``Lepton asymmetries in exclusive b --> s l+ l- decays as a test of the
  %standard model,''
  Phys.\ Lett.\  B, 1998, {\bf 430}, 332
  [arXiv:hep-ph/9803343].
  %%CITATION = PHLTA,B430,332;%%

\bibitem{SM-NP}
%\bibitem{Burdman:1995ks}
  Burdman G,
  %``Testing the standard model in B $\to$ K(*) lepton+ lepton-,''
  Phys.\ Rev.\  D, 1995, {\bf 52}, 6400
  [arXiv:hep-ph/9505352];
  %%CITATION = PHRVA,D52,6400;%%
%\bibitem{Burdman:1998mk}
  Burdman G,
  %``Short distance coefficients and the vanishing of the lepton asymmetry in $B
  %\to$ V $\ell^+$ lepton-,''
  Phys.\ Rev.\  D, 1998, {\bf 57}, 4254
  [arXiv:hep-ph/9710550];
  %%CITATION = PHRVA,D57,4254;%%
%\bibitem{Beneke:2001at}
  Beneke M, Feldmann T and Seidel D,
  %``Systematic approach to exclusive B --> V l+ l-, V gamma decays,''
  Nucl.\ Phys.\  B, 2001, {\bf 612}, 25
  [arXiv:hep-ph/0106067];
  %%CITATION = NUPHA,B612,25;%%
%\bibitem{Feldmann:2002iw}
  Feldmann T and Matias J,
  %``Forward-backward and isospin asymmetry for B --> K* l+ l- decay in the
  %standard model and in supersymmetry,''
  JHEP 2003, {\bf 0301}, 074
  [arXiv:hep-ph/0212158];
  %%CITATION = JHEPA,0301,074;%%
%\bibitem{Kruger:2005ep}
  Kruger F and Matias J,
  %``Probing new physics via the transverse amplitudes of B0 --> K*0 (--> K-
  %pi+) l+ l- at large recoil,''
  Phys.\ Rev.\  D, 2005, {\bf 71}, 094009
  [arXiv:hep-ph/0502060].
  %%CITATION = PHRVA,D71,094009;%%

\bibitem{NPstudies}
%\bibitem{Hewett:1996ct}
 Hewett J L and Wells J D,
  %``Searching for supersymmetry in rare B decays,''
  Phys.\ Rev.\  D, 1997, {\bf 55}, 5549
  [arXiv:hep-ph/9610323];
  %%CITATION = PHRVA,D55,5549;%%
%\bibitem{Ali:1999mm}
 Ali  A, Ball P, Handoko L T and Hiller G,
  %``A Comparative study of the decays $B \to$ ($K$, $K^{*)} \ell^+ \ell^-$ in
  %standard model and supersymmetric theories,''
  Phys.\ Rev.\  D, 2000, {\bf 61}, 074024
  [arXiv:hep-ph/9910221];
  %%CITATION = PHRVA,D61,074024;%%
%\bibitem{Colangelo:2006vm}
  Colangelo P, De Fazio F., Ferrandes R. and Pham T.~N.,
  %``Exclusive B --> K(*) l+ l-, B --> K(*) nu anti-nu and B --> K* gamma
  %transitions in a scenario with a single universal extra dimension,''
  Phys.\ Rev.\  D, 2006, {\bf 73}, 115006
  [arXiv:hep-ph/0604029];
  %%CITATION = PHRVA,D73,115006;%%
%\bibitem{Hovhannisyan:2007pb}
  Hovhannisyan A, HOU W~S and Mahajan N,
  %``B --> K* l+ l- forward-backward asymmetry and new physics,''
  Phys.\ Rev.\  D, 2008, {\bf 77}, 014016
  [arXiv:hep-ph/0701046].
  %%CITATION = PHRVA,D77,014016;%%

\bibitem{OldExp}
%\bibitem{Ishikawa:2006fh}
  Ishikawa A {\it et al.},
  %``Measurement of forward-backward asymmetry and Wilson coefficients in B  -->
  %K* l+ l-,''
  Phys.\ Rev.\ Lett.\ 2006, {\bf 96}, 251801
  [arXiv:hep-ex/0603018].
  %%CITATION = PRLTA,96,251801;%%

\bibitem{Babar:BtoK1}
  Aubert B {\it et al.}  [BABAR Collaboration],
  %``Direct CP, Lepton Flavor and Isospin Asymmetries in the Decays $B \to
  %K^{(*)} \ell^{+} \ell^{-}$,''
  Phys.\ Rev.\ Lett.\ 2009, {\bf 102}, 091803
  [arXiv:0807.4119 [hep-ex]].

\bibitem{Barbar:BtoK2}
  Aubert B {\it et al.}  [BABAR Collaboration],
  %``Angular Distributions in the Decays $B \to K^* \ell^+\ell^-$,''
  Phys.\ Rev.\  D, 2009, {\bf 79}, 031102
  [arXiv:0804.4412 [hep-ex]].
  %%CITATION = PHRVA,D79,031102;%%

\bibitem{FBSexperiment}
J.~T.~Wei {\it et al.}  [BELLE Collaboration],
  %``Measurement of the Differential Branching Fraction and Forward-Backword
  %Asymmetry for B ---> K(*)l+l-,''
  Phys.\ Rev.\ Lett.\ ,2009, {\bf 103}, 171801
  [arXiv:0904.0770 [hep-ex]].

\bibitem{FCNCZpr}
%\bibitem{Langacker:2000ju}
  Langacker P and Plumacher M,
  %``Flavor changing effects in theories with a heavy $Z^\prime$ boson with
  %family nonuniversal couplings,''
  Phys.\ Rev.\  D, 2000, {\bf 62}, 013006
  [arXiv:hep-ph/0001204].
  %%CITATION = PHRVA,D62,013006;%%

\bibitem{ZprPheno1}
%\bibitem{Barger:2003hg}
  Barger V, Chiang C W, Langacker P and Lee H S,
  %``$Z^\prime$ mediated flavor changing neutral currents in $B$ meson decays,''
  Phys.\ Lett.\  B, 2004, {\bf 580}, 186
  [arXiv:hep-ph/0310073];
  %%CITATION = PHLTA,B580,186;%%
%\bibitem{Barger:2004qc}
  Barger V, C.~W.~Chiang, J.~Jiang and P.~Langacker,
  %``$B_s - \bar{B}_s$ mixing in $Z^\prime$ models with flavor-changing neutral
  %currents,''
  Phys.\ Lett.\  B, 2004 {\bf 596}, 229
  [arXiv:hep-ph/0405108];
  %%CITATION = PHLTA,B596,229;%%
%\bibitem{Barger:2004hn}
  Barger V, C.~W.~Chiang, P.~Langacker and H.~S.~Lee,
  %``Solution to the B --> pi K puzzle in a flavor-changing Z' model,''
  Phys.\ Lett.\  B, 2004, {\bf 598}, 218
  [arXiv:hep-ph/0406126];
  %%CITATION = PHLTA,B598,218;%%
%\bibitem{Arhrib:2006sg}
  Arhrib A, Cheung K, C.~W.~Chiang and T.~C.~Yuan,
  %``Single top-quark production in flavor-changing Z' models,''
  Phys.\ Rev.\  D, 2006, {\bf 73}, 075015
  [arXiv:hep-ph/0602175];
  %%CITATION = PHRVA,D73,075015;%%
%\bibitem{Cheung:2006tm}
  Cheung K, Chiang C~W, Deshpande N~G and Jiang J,
  %``Constraints on flavor-changing Z' models by B/s mixing, Z' production, and
  %B/s --> mu+ mu-,''
  Phys.\ Lett.\  B, 2007, {\bf 652}, 285
  [arXiv:hep-ph/0604223];
  %%CITATION = PHLTA,B652,285;%%
%\bibitem{Chiang:2006we}
  Chiang C W, Deshpande N G and Jiang J,
  %``Flavor changing effects in family nonuniversal $Z^\prime$ models,''
  JHEP, 2006, {\bf 0608}, 075
  [arXiv:hep-ph/0606122].
  %%CITATION = JHEPA,0608,075;%%

\bibitem{ZprPheno2}
%\bibitem{He:2004it}
  HE X G and Valencia G,
  %``K+ --> pi+ nu anti-nu and FCNC from non-universal Z' bosons,''
  Phys.\ Rev.\  D, 2004, {\bf 70}, 053003
  [arXiv:hep-ph/0404229];
  %%CITATION = PHRVA,D70,053003;%%
%\bibitem{He:2007iu}
  %``$D$ - $\bar{D}$ mixing constraints on FCNC with a non-universal
  %$Z^\prime$,''
  Phys.\ Lett.\  B, 2007, {\bf 651}, 135
  [arXiv:hep-ph/0703270].
  %%CITATION = PHLTA,B651,135;%%

\bibitem{ZprPheno3}
%\bibitem{Barger:2009eq}
  Barger V, Everett L, Jiang J, Langacker P, Liu T and Wagner C,
  %``Family Non-universal $U(1)^\prime$ Gauge Symmetries and $b\to s$
  %Transitions,''
  Phys.\ Rev.\  D, 2009, {\bf 80}, 055008
  [arXiv:0902.4507 [hep-ph]];
  %%CITATION = PHRVA,D80,055008;%%
%\bibitem{Barger:2009qs}
  %``$b \to s$ Transitions in Family-dependent $U(1)^\prime$ Models,''
  arXiv:0906.3745 [hep-ph].
  %%CITATION = ARXIV:0906.3745;%%

\bibitem{ZprPheno4}
%\bibitem{Chen:2008za}
  CHEN S L and Okada N,
  %``Flavorful $Z^\prime$ signatures at LHC and ILC,''
  Phys.\ Lett.\  B, 2008, {\bf 669}, 34
  [arXiv:0808.0331 [hep-ph]].
  %%CITATION = PHLTA,B669,34;%%

\bibitem{ZprPheno5}
 CHANG Q, LI X Q and YANG Y D,
  %``Constraints on the nonuniversal $Z^\prime$ couplings from B\to\pi K, \pi
  %K^{\ast} and \rho K Decays,''
  JHEP 2009, {\bf 0905}, 056
  [arXiv:0903.0275 [hep-ph]];
  %``Family Non-universal Z^\prime effects on \bar{B}_q-B_q$ mixing, B\to X_s
  %\mu^+\mu^- and B_s\to \mu^+\mu^- Decays,''
  arXiv:0907.4408 [hep-ph].
\bibitem{PQCD}
%\bibitem{Keum:2000ph}
  Keum Y Y, Li H n and Sanda A I,
  %``Fat penguins and imaginary penguins in perturbative QCD,''
  Phys.\ Lett.\  B, 2001, {\bf 504}, 6
  [arXiv:hep-ph/0004004];
  %%CITATION = PHLTA,B504,6;%%
  %%CITATION = PHRVA,D63,074006;%%
LU CD, Ukai K,  YANG M Z, Phys. Rev.  D, 2001, {\bf 63},  074009
e-Print:
hep-ph/0004213; %\bibitem{Keum:2000ms}
  Keum Y  and Li H~n,
  %``Nonleptonic charmless B decays: Factorization vs. perturbative QCD,''
  Phys.\ Rev.\  D, 2001, {\bf 63}, 074006
  [arXiv:hep-ph/0006001];
  LU C~D,  YANG M Z, Eur. Phys. J. C, 2002, {\bf  23}, 275-287
 e-Print: hep-ph/0011238


\bibitem{formfactorsQCDSR}
  Ball P and Braun V M,
  %``Exclusive semileptonic and rare B meson decays in {QCD},''
  Phys.\ Rev.\  D, 1998, {\bf 58}, 094016
  [arXiv:hep-ph/9805422].



\bibitem{BtoKbyAli}
 Ali A, Ball P, Handoko L~T, and Hiller G, Phys.\ Rev.\ D, 2000, {\bf
 61}, 074024.


\bibitem{Buchalla:1995vs}
  Buchalla G, Buras A J and Lautenbacher M E,
  %``Weak Decays Beyond Leading Logarithms,''
  Rev.\ Mod.\ Phys.\   1996, {\bf 68}, 1125
  [arXiv:hep-ph/9512380].
  %%CITATION = RMPHA,68,1125;%%




\bibitem{ypert}
  Buras A J and Munz M,
  %``Effective Hamiltonian for B $\to$ X(s) e+ e- beyond leading logarithms in
  %the NDR and HV schemes,''
  Phys.\ Rev.\  D, 1995, {\bf 52}, 186
  [arXiv:hep-ph/9501281].

\bibitem{BtoK1}
  LI R H , LU C D and WANG W,
  %``Branching Ratios, Forward-backward Asymmetry and Angular Distributions of
  %$B\to K_1\ell^+\ell^-$ Decays,''
  Phys.\ Rev.\  D 2009, {\bf 79}, 094024
  [arXiv:0902.3291 [hep-ph]].
  %%CITATION = PHRVA,D79,094024;%%

\bibitem{rhosbL}
 Cheung K, Chiang C W, Deshpande N G and Jiang J, Phys. Lett.
 B, 2007, {\bf 652},285

\bibitem{rhosbL2}
  V.~Barger, L.~Everett, J.~Jiang, P.~Langacker, T.~Liu and C.~Wagner,
  %``Family Non-universal U(1)-prime Gauge Symmetries and b ---> s
  %Transitions,''
  Phys.\ Rev.\  D, 2009, {\bf 80}, 055008[arXiv:0902.4507 [hep-ph]];
  arXiv:0906.3745[hep-ph]; V.~Barger, L.~L.~Everett, J.~Jiang, P.~Langacker, T.~Liu and C.~E.~M.~Wagner,
  %``b ---> s Transitions in Family-dependent U(1)-prime Models,''
  JHEP, 2009, {\bf 0912}, 048[arXiv:0906.3745 [hep-ph]]

\bibitem{Aaltonen:2007kv}
  Aaltonen T {\it et al.}  [CDF Collaboration],
  %``Search for $B^0_{s} \to \mu^{+} \mu^{-}$ and $B^0_{d} \to \mu^{+} \mu^{-}$
  %decays with $2fb^{-1}$ of $p \bar{p}$ collisions,''
  Phys.\ Rev.\ Lett.\ 2008, {\bf 100}, 101802
  [arXiv:0712.1708 [hep-ex]].

\bibitem{Abazov:2007iy}
  Abazov V M{\it et al.}  [D0 Collaboration],
  %``Search for $B_s \to \mu^{+} \mu^{-}$ at D0,''
  Phys.\ Rev.\  D 2007, {\bf 76}, 092001
  [arXiv:0707.3997 [hep-ex]].

\end{thebibliography}
\end{document}